\documentclass[prl,twocolumn,showpacs,showkeys,preprintnumbers,superscriptaddress,floatfix]{revtex4-1}
\usepackage[T1]{fontenc}
\usepackage[utf8]{inputenc}
\inputencoding{latin1}
\usepackage{amsbsy}
\usepackage{amstext}
\usepackage{graphicx}

\makeatletter

\usepackage[english]{babel}
\usepackage{amsmath}
\usepackage{amsfonts}
\usepackage{color}
\usepackage{amssymb}
\usepackage{array}
\usepackage{makeidx}
\usepackage{float}
\usepackage{multirow}
\usepackage{times}

\usepackage{wasysym}
\newcommand{\BCS}{BaCoS$_{2}$}
\newcommand{\BCNS}{BaCo$_{1-x}$Ni$_{x}$S$_{2}$}

\usepackage[svgnames]{xcolor}
\newif\ifshowcomments\showcommentstrue
\def\beq{\begin{equation}}
\def\eeq{\end{equation}}

\makeatother

\usepackage{lineno}

\begin{document}

\title{Two-dimensional fluctuations and competing phases in the stripe-like antiferromagnet \BCS}

\author{Haneen Abushammala}
\affiliation{IMPMC, Sorbonne Universit\'e, CNRS and MNHN, 4, place Jussieu, F-75005 Paris, France}

\affiliation{Institute for Experimental Physics IV, Ruhr-Universit\"at Bochum, 44801 Bochum, Germany}

\author{Benjamin Lenz}
\affiliation{IMPMC, Sorbonne Universit\'e, CNRS and MNHN, 4, place Jussieu, F-75005 Paris, France}

\author{Benoit Baptiste}
\affiliation{IMPMC, Sorbonne Universit\'e, CNRS and MNHN, 4, place Jussieu, F-75005 Paris, France}

\author{David Santos-Cottin}
\affiliation{Department of Physics, University of Fribourg, 1700 Fribourg, Switzerland}

\author{Pierre Toulemonde}
\affiliation{CNRS, Universit\'e Grenoble Alpes, Institut N\'eel, 38000, Grenoble, France}

\author{Michele Casula}
\affiliation{IMPMC, Sorbonne Universit\'e, CNRS and MNHN, 4, place Jussieu, F-75005 Paris, France}

\author{Yannick Klein}
\affiliation{IMPMC, Sorbonne Universit\'e, CNRS and MNHN, 4, place Jussieu, F-75005 Paris, France}

\author{Andrea Gauzzi}
\affiliation{IMPMC, Sorbonne Universit\'e, CNRS and MNHN, 4, place Jussieu, F-75005 Paris, France}

\begin{abstract}
By means of a x-ray diffraction, magnetic susceptibility and specific heat study, we investigate the interplay between orthorhombic distortion and stripe-like antiferromagnetic (AFM) order at $T_N=290$ K in the Mott insulator \BCS. The data analysis give evidence of a purely electronic AFM transition with no participation of the lattice. The observation of large thermal fluctuations in the vicinity of $T_N$ and of a Schottky anomaly are ascribed to the quasi-two dimensional character of the magnetic structure and to the existence of competing ground states within a minute $\sim$1 meV energy range that differ in the orbital and spin configurations of the Co ions. This scenario suggests that the stripe-like AFM order results from a spontaneous symmetry breaking of the geometrically frustrated pristine tetragonal phase, which offers an ideal playground to study the driving force of multi-orbital Mott transitions without the participation of the lattice.    
\end{abstract}

\maketitle

The single-band Hubbard model has been widely employed to describe the Mott metal-insulator transition (MIT) in a variety of correlated electron systems \cite{ima98}. The model has been very successful to account for the MIT phase diagram as a function of electronic doping, $n$, and of the $U/t$ ratio, where $U$ and $t$ are the onsite Coulomb repulsion and the hopping energy between nearest neighbour sites, respectively. In fact, the applicability of the model to real systems is limited because most Mott materials display a strong interplay between the above electronic parameters and the lattice, \textit{e.g.} structural distortions, strain fields and disorder, thus preventing a clear description of the stability conditions of the metallic and insulating phases \cite{anderson1972effect,lee85,di2017disorder}. This is indeed the case of well-studied prototype Mott systems, such as V$_2$O$_3$ \cite{V2O31973vv} and BaVS$_3$ \cite{foury2012groundBaVS3}.

Here, we consider \BCNS\ (BCNS) \cite{martinson1993metal} as a model Mott system where the driving force of the MIT is purely electronic and no coupling with the lattice occurs at the transition, as originally imagined by Mott. As expected following Mott's criterion and similarly to other typical Mott systems, the ground state of the pristine unsubstituted \BCS\ phase is insulating and antiferromagnetic (AFM), while the metallic phase is stabilized upon doping above a critical chemical substitution value $x_{cr} =$ 0.22 \cite{takeda1995transport}. Unusual is the fact that, in spite of a modest orthorhombic distortion of the $ab$-plane ($a=$ 6.488 \AA, $b=$ 6.439 \AA~ \cite{grey1970crystal}), the AFM structure of \BCS\ is stripe-like, i.e. the moments of the Co$^{2+}$ ions are AFM- and FM-ordered along the $a$- and $b$-directions, respectively, as indicated by previous neutron diffraction studies \cite{kod96,man97}.

In the present work, we investigate the stability conditions of such unusual stripe-like AFM order. To do so, we first elucidate the relationship between orthorhombic distortion and AFM order in high-quality \BCS\ powders and single crystals. Second, we study the structural and magnetic properties of a metastable \BCS\ phase with reduced orthorhombic distortion obtained by quenching the pristine \BCS\ sample under high-pressure from high temperature. We argue that the present study may elucidate the role of the magnetic order on the MIT in \BCNS\ and also the interplay between spontaneous symmetry breaking, orbital and spin degrees of freedom in other multi-orbital strongly correlated electron systems, such as Fe-based superconductors \cite{gla15,wan15}.   

We prepared single-phase \BCS\ powder samples following the solid-state synthesis route previously reported \cite{gel96,sny94,man97,fis99}. In brief, stoichiometric quantities of BaS (99.998$\%$, sigma Aldrich), Co (99.999$\%$, sigma Aldrich) and S (99.98$\%$, sigma Aldrich) were ground and pressed in a glove-box under Ar atmosphere. The pellets were then loaded in a graphite crucible, sealed under vacuum in quartz ampoules, heated up to 950$^{\circ}$C for 72h and finally quenched in a bath of liquid nitrogen to avoid phase decomposition. In order to investigate the stability of the \BCS\ phase, some of the as-prepared powders have been submitted to a second heat treatment at high pressure (HP) in a belt-type of press at 4 GPa and 940$^{\circ}$C for 1 hour and then quenched to room temperature by holding the pressure constant \cite{bra07}.

We studied the crystal structure of the as-prepared samples by X-ray diffraction in the Bragg-Brentano reflection geometry using a Panalytical XpertPro MPD 2-circle diffractometer equipped with a Co K$_{\alpha}$ source at room temperature and at low-temperature in the 100-400 K range using an Anton Paar HTK 450 temperature chamber. The the diffraction data were refined using the Rietveld method implemented in the FullProf software \cite{rod91}. Temperature-dependent magnetic susceptibility, $\chi$, and constant pressure specific heat, $C_P$, measurements were performed within the 2-400 K range using a commercial Quantum Design SQUID magnetometer and a commercial physical properties measurement system (PPMS), respectively. The specific heat was measured using a standard 2$\tau$-relaxation method.

\begin{figure}[ht]
\centering
\includegraphics[width=\columnwidth]{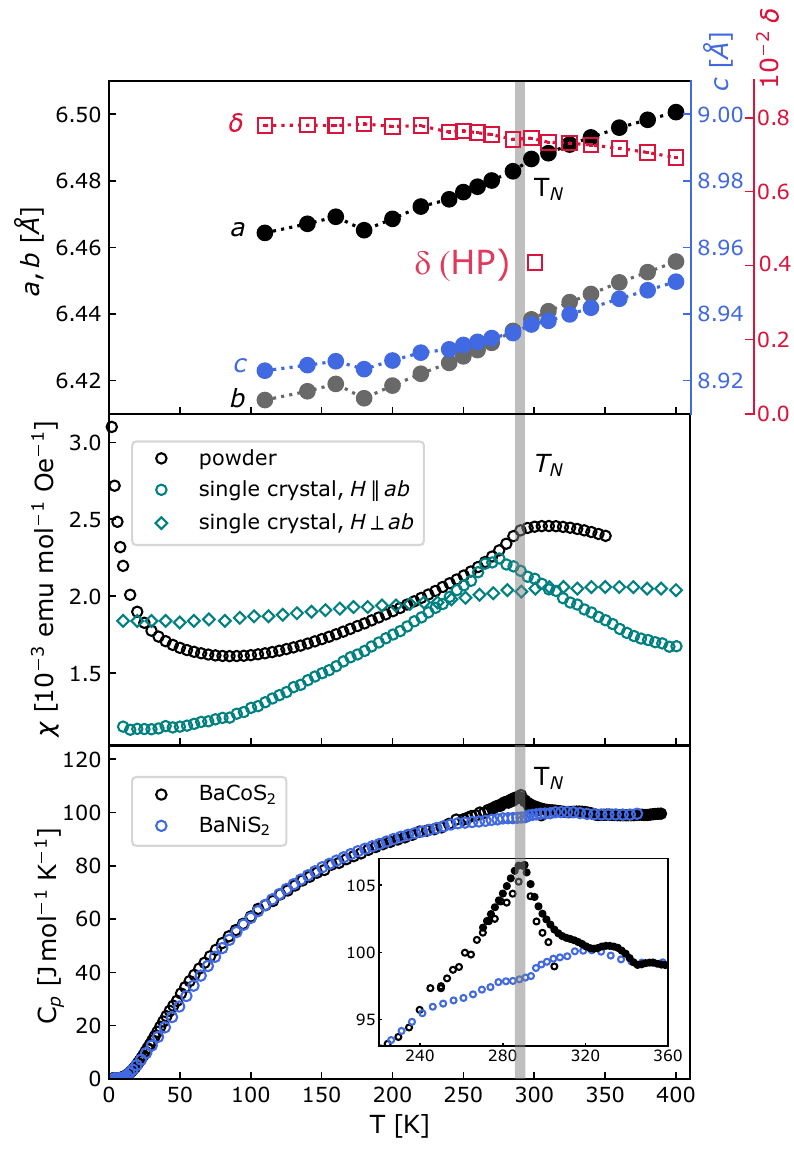}
\caption{Top panel: temperature evolution of the $Cmma$ cell parameters and of the orthorhombic distortion, $\delta$, of pristine \BCS\ powders prepared under ambient pressure. Middle and bottom panels: the same as above for the zero field cooling (ZFC) magnetic susceptibility and of the specific heat. In the top panel, the room temperature $\delta$ value of a powder sample quenched at high pressure (HP) is also shown for comparison. In the middle panel, the susceptibility curve of a single crystal with field parallel and perpendicular to the $ab$ plane is shown for comparison, which shows that the ordered magnetic moments lie in the $ab$-plane. In the bottom panel, open and full black circles refer to two separate measurements at low and high temperatures. For comparison, the data of a nonmagnetic BaNiS$_2$ powder sample is also reported, which enables us to extract the electronic contribution at the AFM transition, $T_N$, as explained in the text. In the inset, we show a detail of the transition.}
\label{fig:TvsorthochiHC}
\end{figure}

We shall first illustrate the results obtained on the pristine powders. Our analysis of the x-ray diffraction data indicates a phase purity of 95\% or better. At all temperatures studied, the diffraction patterns have been successfully refined using the Rietveld method. Our analysis indicates that the orthorhombic $Cmma$ symmetry explains well the data not only at room temperature, as previously reported \cite{gel96,sny94}, but in the whole 100-400 K range studied, as indicated by reliability $R$-Bragg factors less than 10\%. Specifically, contrary to previous reports \cite{bae94,JT2020,gel96}, we find no indications of structural distortion or modulation that would lower the crystal symmetry. In Table I of the SM, we report the results of representative Rietveld refinements of the room temperature data taken on both, pristine and HP/HT sample. In Figure \ref{fig:TvsorthochiHC}, we plot the lattice parameters as a function of temperature. Within the experimental uncertainty of $\pm 0.5$ pm or better, we see no anomaly of any of these parameters at the AFM ordering temperature, $T_N$. Also, the magnitude of the orthorhombic distortion, defined as $\delta = 2(a-b)/(a+b)$, remains nearly constant over the whole temperature range studied. This observation clearly indicates the absence of a significant interplay between lattice and magnetic order and that the AFM transition is driven solely by the electronic degrees of freedom.

The $\chi$ and $C_P$ behavior of the same \BCS\ powder sample is shown in the middle and bottom panels of Figure \ref{fig:TvsorthochiHC}, respectively. Both curves display a clear anomaly at $T_N$, which we define as the specific heat peak at $\sim 290$ K. Contrary to previous reports, we do not define $T_N$ as the temperature of the cusp-like anomaly in the $\chi$ curve. Indeed, this anomaly does not necessarily reflect the onset of a long-range AFM order but rather an abrupt suppression of the spin susceptibility, as previously found in other Mott systems, such as the aforementioned BaVS$_3$ \cite{mih00}, where no long-range AFM order occurs at the metal-insulator transition. We shall confirm later the validity of this observation in the present case. Notable is the large fluctuation region around $T_N$ suggesting strong short-range fluctuations and a reduced dimensionality of the transition. This observation suggests a scenario of quasi-twodimensional magnetic order where the magnetic moments lie within the $ab$-plane, consistent with the aforementioned neutron diffraction studies \cite{kod96,man97} and by the strong anisotropy of the susceptibility measured on an oriented single crystal, shown in Figure \ref{fig:TvsorthochiHC}. Namely, the model of two-dimensional ordering of the moments within the $ab$-plane is corroborated by the fact that the characteristic cusp at $T_N$ in the ZFC curve is seen only for fields parallel to the $ab$-plane, while the susceptibility curve for field perpendicular to the plane is almost flat.

In Figure \ref{fig:TvsorthochiHC}, we also plot the $C_P(T)$ curve of a BaNiS$_2$ powder sample, which enables us to extract the electronic contribution to the specific heat associated with the AFM transition. Indeed, BaNiS$_2$ is an isostructural and nonmagnetic phase of \BCS\ and the mass difference between Co and Ni is small, so we expect that the lattice contribution to the specific heat is unchanged. As expected, the two curves are almost identical at all temperatures, except in the transition region. Therefore, by subtracting the BaNiS$_2$ curve from the \BCS\ one, we numerically compute the change of entropy associated with the transition as the integral $\Delta S_{\rm mag}=\int_{T_1}^{T_2}\frac{\Delta C_P(T)}{T} dT$, where $\Delta C_P(T)$ is the difference between the $C_P/T$ curves of the two samples and $T_1 - T_2$ is a sufficiently large temperature interval, 230-360 K, encompassing the transition region. We obtain a very small value $\Delta S_{\rm mag}=0.6$ J mol$^{-1}$ K$^{-1}$, as compared to the value $k_B {\rm ln} 2 = 5.76$ J mol$^{-1}$ K$^{-1}$ expected for an ideal AFM structure formed by localized moments. This discrepancy suggests a picture of delocalized moments, consistent with the reduced ionicity of transition metal sulfides, or the existence of large AFM fluctuations well above $T_N$.
\begin{figure}[ht]
\centering
\includegraphics[width=\columnwidth]{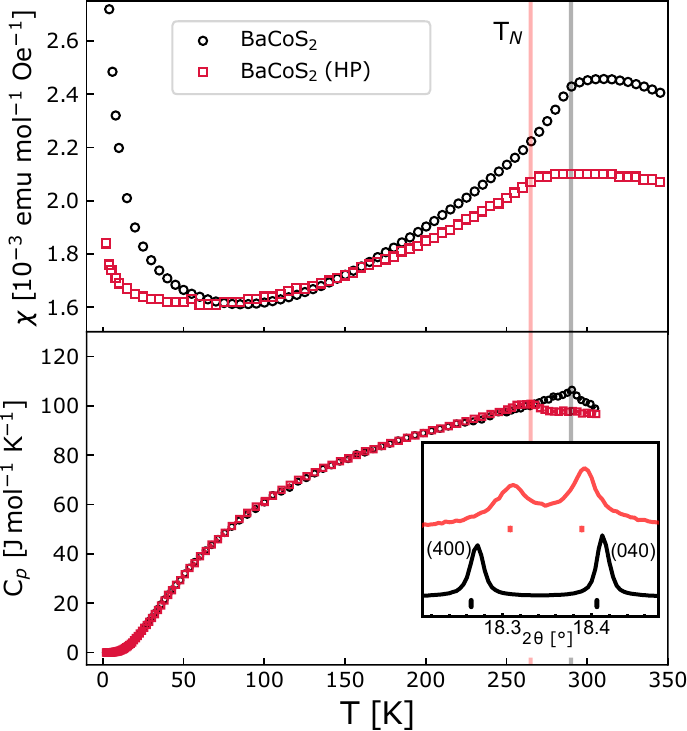}
\caption{Temperature dependence of the ZFC magnetic susceptibility and of the specific heat of the pristine (black symbols) and HP quenched (red symbols) \BCS\ powders. The vertical bar indicates the AFM transition at $T_N$. Inset: detail of the x-ray diffraction patterns of the two samples showing the reduced splitting of the (400) and (040) reflections and hence a reduced orthorhombic distortion, $\delta$, of the latter sample (see also the top panel of Fig. \ref{fig:TvsorthochiHC}).}
\label{fig:orthovscpchi}
\end{figure}

\begin{figure}[ht]
\centering
\includegraphics[width=\columnwidth]{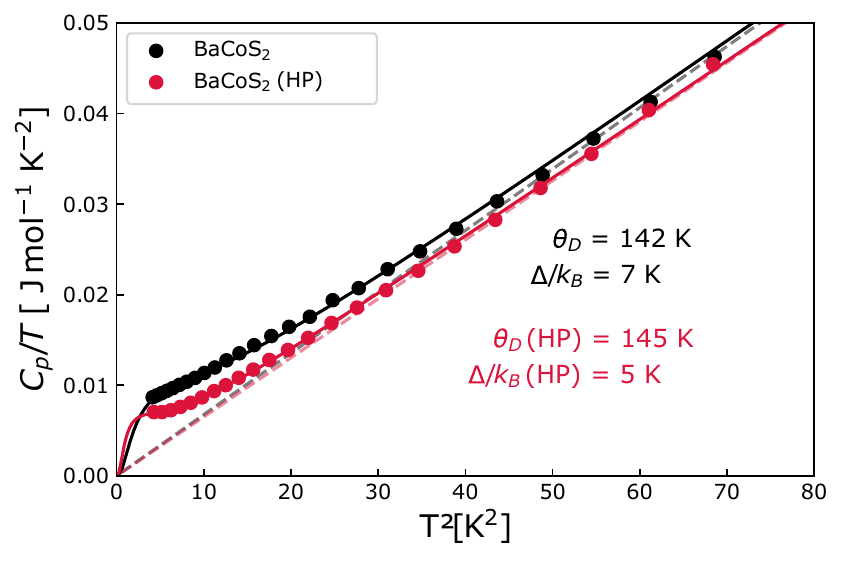}
\caption{Low-temperature behavior of the specific heat of the pristine and HP quenched \BCS\ powders of Fig. \ref{fig:orthovscpchi}. Broken lines are linear fits above 6 K, which yields $\gamma=0$ for both samples and the Debye temperature values, $\theta_D$, in the legend. Solid lines are fits including the two-level Schottky term of Eq. (1), which yields the value of energy separation, $\Delta$, between the two levels in the legend.}
\label{fig:Schottky}
\end{figure}

We now illustrate the results on the HP quenched sample. The analysis of the X-ray diffraction pattern indicates that the HP quenching of the pristine powders does not alter either phase purity or the crystalline quality, as indicated by a similar quality of the refined structure (see Table I in the SM and in Figure \ref{fig:orthovscpchi}). In the HP sample the thermal parameters of the Ba and Co sites are larger, which suggests a larger disorder. Notable is the difference in the lattice parameters and in the apical Co-S1 distance between the two samples. Namely, the orthorhombic distortion of the HP sample is about half of that of the pristine sample (see Figure  \ref{fig:TvsorthochiHC}) and the apical Co-S1 distance is significantly shorter by 0.012 \AA. 

The above structural differences are concomitant to remarkably different physical properties. In Figure \ref{fig:TvsorthochiHC}, we note a sizable $\sim$30 K reduction of $T_N$ accompanied by a dramatic broadening of the $C_P$ peak and of the $\chi$ cusp-like anomaly. The broadenings are so large that one may question whether the long-range AFM order is still present in the HP sample. We obtain further insights into the stability of the AFM phase and on the effects of the HP quenching by investigating the low-temperature behavior of the specific heat of both pristine and quenched samples. Within a conventional Debye model, at sufficiently low temperatures, one expects a linear $C_P/T$ vs. $T^2$ dependence, $C_P/T = \gamma + \beta T^2$, where $\beta = 12 \pi^4 R / 5 \theta^{3}_{D}$, $R$ is the gas constant, $\theta_D$ is the Debye temperature and the Sommerfeld coefficient, $\gamma$, is expected to vanish because \BCS\ is an insulator. Consistent with this expectation, both experimental curves of Figure \ref{fig:Schottky} exhibit a marked linear dependence that extrapolates to the origin. Above $T \sim 6$ K, the two curves are identical within the experimental uncertainty, which corroborates the previous observation that the structural (and hence dynamical) properties are very similar. At lower temperatures, one notes an upturn in both samples, the signature of an additional entropy contribution. While the onset temperatures of the upturn are comparable in the two samples, the upturn is more pronounced in the HP sample.

Following previous reports on a variety of systems displaying competing ground states, such as spin glasses \cite{hei87}, magnetic clusters \cite{ana12} and frustrated magnetic compounds \cite{raj92}, we attribute this anomaly to a Schottky term that appears in the presence of two or more nearly-degenerate states within an energy range of the order of the thermal energy. In order to verify this possibility, we consider a minimal two-state model described by the expression:

\begin{equation}\label{eq:1}
\frac{C(T)}{T} = \gamma + \beta T^2 + 4\ R\ \frac{z^{2}}{T}\frac{e^{z}}{(1+e^{z})^{2}} 
\end{equation}

where the third (Schottky) term is expressed as a function of the adimensional variable $z=\Delta/k_BT$ and $\Delta$ is the energy separation between the two states. The Schottky term produces a broad hump in the $C_P/T$ vs $T^2$ plots at temperatures $T \sim \Delta /k_B$, which corresponds to a maximum increase of entropy between the zero-temperature configuration, where only the ground state is populated, and the high-temperature configuration, where the two states are equally populated.

As seen in Figure \ref{fig:Schottky}, the experimental $C_P(T)$ curves of both, pristine and HP samples, are well explained by Eq. (1). A straightforward fit yields $\gamma =0$ for both samples, as expected for an insulator, and very similar Debye temperatures, $\theta_D$ = 142 and 145 K for the pristine and HP sample, respectively, consistent with the observation that a reduced orthorhombic distortion in the HP quenched sample would not alter significantly the lattice contribution to the specific heat. Though, we obtain quite different $\Delta$ values, namely $\Delta/k_B =$ 7.2 and 5 K, for the pristine and HP sample, respectively. This suggests the existence of nearly-degenerate states whose energy separation is smaller in the HP sample. In both curves, the upturn corresponds to the high-temperature tail of the hump described by Eq. (1), the broad maximum of the hump is not visible in the data, for it would occur below the lowest temperature measured of 2 K. 

\begin{figure}[ht]
\centering
\includegraphics[width=\columnwidth]{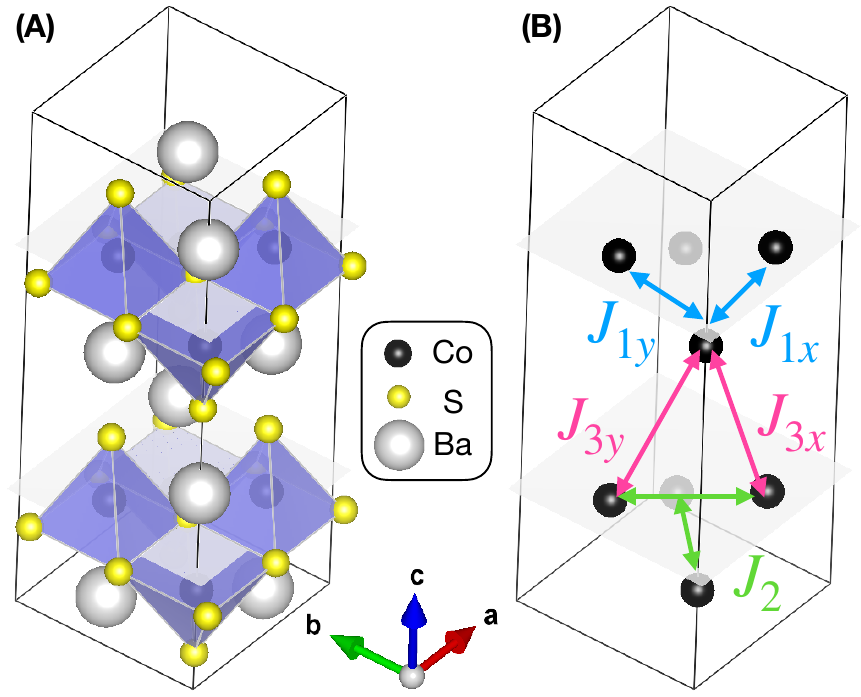}
\caption{
(A) Crystal structure of \BCS\ formed by puckered layers of CoS$_6$ tetrahedra separated by Ba atoms. (B) Schematics of the minimal $J_1-J_2-J_3$ Heisenberg spin model proposed in Ref.~\onlinecite{lenz23}. 
}
\label{fig:J}
\end{figure}

Hereafter, we should attempt to provide a qualitative explanation of the above finding supported by an extensive \textit{ab initio} study presented in a parallel publication by some of us \cite{lenz23}. We first recall that the magnetic stripe order in \BCS\ had previously been connected to frustrated quasi-2D models like the $J_1-J_2$ Heisenberg model \cite{Mila91,san16b}, which explains this order for $J_2\gg J_1$, but fails to explain the surprisingly high N\'eel temperature. The origin of this frustration is apparent in Figure \ref{fig:J}, where we highlight the two different exchange constants, $J_1$ and $J_2$, involving nearest and next-nearest neighbors respectively of the Co ions in the nearly square lattice of BaCoS$_2$.
In the absence of lattice distortions or orbital anisotropy, this geometrical frustration prevents any long-range AFM order.
The above study \onlinecite{lenz23} shows that this order is stabilized by an order-disorder transition mechanism described by a frustrated $J_1-J_2-J_3$ Heisenberg Hamiltonian that includes the out-of-plane exchange interaction, $J_3$ (see Figure ~\ref{fig:J}). 
The parameter regime of the model is motivated by density functional theory (DFT)+U calculations. 
The calculations support the existence of several low-lying magnetic states within a small energy window of $20$ meV with respect to the stripe AFM ground state. 
Consistent with our analysis of the low-temperature specific heat data, the energy of the first excited state is predicted to be only a few kelvin higher than the ground state. 

In agreement with previous DFT+U and dynamical mean-field theory calculations \cite{zainullina2011,san18}, such competing states correspond to different occupations of the $d$-orbitals and to different spin configurations of the Co ions, thus breaking the isotropy of the $ab$-plane and stabilizing the magnetic stripe order in the plane. The existence of quasi-degenerate states probed experimentally by the present low-temperature specific heat measurement accounts well for the observation of large thermal fluctuations in the specific heat transition. Indeed, the vicinity of competing states enhances the fluctuations. The proposed scenario suggests that, at temperatures sufficiently high above $T_N$, the AFM order melts into a mixture of other spin 
states rather than into a paramagnetic state lying at much higher energy. Within this scenario, the reduced orthorhombic distortion forced by the HP quenching enhances the latent geometrical frustration, thus leading to a more pronounced near degeneracy of competing states. As a consequence, the AFM order is weakened ($T_N$ is lowered) and the thermodynamic fluctuations in the transition region are further enhanced. This phenomenology is again consistent with the scenario of an Ising-like transition driven by an order-disorder transition mechanism proposed in Ref.~\onlinecite{lenz23}.

In conclusion, we have investigated the stability of the peculiar stripe-like AFM ground state of the multiorbital Mott insulator \BCS\ by means of a combined structural, magnetic and specific heat study on the stable \BCS\ phase and on a metastable one quenched under high-pressure. A first finding is the great stability of the orthorhombic $Cmma$ phase within a broad 100-400 K range with no anomaly at the AFM transition, which points at a purely electronic origin of the transition with no participation of the lattice. Second, the low-temperature behavior of the specific heat unveils that the AFM ground state competes with other low-lying states within a minute $\sim$1 meV energy range. This result is explained by the latent geometric frustration of the magnetic exchange interactions in the tetragonal phase and accounts for a dramatic enhancement of thermodynamic fluctuations at the transition. The emerging scenario is that the stripe-like AFM structure of \BCS\ results from a spontaneous breaking of the pristine tetragonal ($C_4$) symmetry which removes frustration by stabilizing an orbital and spin configuration with orthorhombic ($C_2$) symmetry. Owing to the absence of a coupling of the electronic degrees of freedom with the lattice, \BCS\ appears to be a model system to unveil the complex interplay between orbital and spin degrees of freedom in multiorbital Mott transitions.

\begin{acknowledgments}
We acknowledge Michele Fabrizio for fruitful discussions, the low-temperature physical measurements (MPBT) platform of Sorbonne Universit\'e for assistance in the magnetization and specific heat measurements, M. Legendre for help in the HP sample synthesis, Campus France and the MATISSE Labex for financial support and the French 'Grand Equipement National de Calcul Intensif' (GENCI) under Project Numbers A0110912043 and A0110906493 for allocation of computation time.
\end{acknowledgments}

\bibliography{biblio_BaCoS2}

\end{document}